\documentclass[12pt,titlepage]{article}
\usepackage{amssymb}
\usepackage[dvips]{graphicx}
\usepackage{subfigure}

\begin{document}

\title{Trapping of a particle in a short-range harmonic potential well}
\author{L.B. Castro and A.S. de Castro\thanks{%
Corresponding author: A.S. de Castro, castro@pq.cnpq.br, tel.
number:+551231232800, fax number:+551231232840.} \\
\\
UNESP - Campus de Guaratinguet\'{a}\\
Departamento de F\'{\i}sica e Qu\'{\i}mica\\
12516-410 Guaratinguet\'{a} SP - Brazil\\
}
\date{}
\maketitle

\begin{abstract}
Eigenstates of a particle in a localized and unconfined harmonic potential
well are investigated. Effects due to the variation of the potential
parameters as well as certain results from asymptotic expansions are
discussed. \newline
\end{abstract}

\section{Introduction}

Systems confined have received considerable attention in quantum mechanics.
The hydrogen atom confined in a spherical enclosure was first analyzed in
1937 \cite{mic}, the restricted rotator in 1940 \cite{som1} and the harmonic
oscillator in 1943 \cite{cha}. Since then the one-dimensional oscillator
immersed in an infinite square potential has received some attention \cite%
{aul}-\cite{vaw}. It should also be mentioned that two different cases for a
sort of one-dimensional half-oscillator have also been reported, one of them
bound by an infinite wall \cite{dea}, \cite{mei} and the other one by a
finite step potential \cite{mei}. Recently, the D-dimensional confined
harmonic oscillator appeared in the literature \cite{mon1}-\cite{mon2}.

The square well potential is an unconfined potential with vertical walls
that has been used to model band structure in solids \cite{kro} and
semiconductor heterostructures \cite{esa}-\cite{red}. The use of a quantum
well with sloping sides might be of interest to refine those models. As a
matter of fact, a sort of localized triangular potential has been an item of
recent practical \cite{cha2}-\cite{ban} and theoretical \cite{lui}-\cite%
{castro} investigations.

In this paper we consider the bound-state problem for a particle immersed in
an one-dimensional harmonic potential which vanishes outside a finite
region. To the best of our knowledge this sort of trapping has never been
solved. By using confluent hypergeometric functions the process of solving
the Schr\"{o}dinger equation for the eigenenergies is transmuted into the
simpler and more efficient process of solving a transcendental equation.
Such as for the well-known square potential, a graphical method provides
some qualitative conclusions about the spectrum of this short-range
potential well. Approximate analytical results for the special cases of
low-lying states and high-lying states are obtained with the help of
asymptotic representations and limiting forms for the confluent
hypergeometric function. It is shown that the localized harmonic potential
yields the full harmonic potential and the square well potential as limiting
cases. Although the quantization condition has no closed form expressions in
terms of simpler functions, the exact computation of the allowed
eigenenergies can be done easily with a root-finding procedure of a symbolic
algebra program. Proceeding in this way, the whole bound-state spectrum is
found. Nevertheless, our purpose is to investigate the basic nature of the
phenomena without entering into the details involving specific applications.
In other words, the aim of this paper is to explore a simple system which
can be of help to see more clearly what is going on into the details of a
more specialized and complex circumstance such as that one in Ref. \cite{cru}%
.

\section{The particle in a short-range harmonic potential well}

Let us write the short-range harmonic potential well as
\begin{eqnarray}
V(x) &=&\frac{1}{2}m\omega ^{2}\left( x^{2}-rL^{2}\right) \left[ \theta
\left( x+L\right) -\theta \left( x-L\right) \right]  \nonumber \\
&&  \nonumber \\
&=&\left\{
\begin{array}{c}
\frac{1}{2}m\omega ^{2}\left( x^{2}-rL^{2}\right) \\
\\
0%
\end{array}%
\begin{array}{c}
{\textrm{for }}|x|<L \\
\\
{\textrm{for }}|x|>L%
\end{array}%
\right.  \label{VV}
\end{eqnarray}%
where $\theta \left( x\right) $ is the Heaviside function, $2L$ is the range
of the potential, $m$ is the mass of the particle and $\omega $ is the
classical frequency of the oscillator. The parameter $r=V(0)/[V(0)-V(L)]$
characterizes four different profiles for the potential as illustrated in
Fig. 1. This potential admits scattering states (with $E>0$) and bound
states (with $V(0)<E<0$ and $r>0$). In what follows we will consider the
bound-state problem.

Let us introduce the new variable%
\begin{equation}
z=\alpha \,x,\quad \alpha =\sqrt{\frac{2m\omega }{\hslash }}  \label{w}
\end{equation}%
so that, for $|x|<L$, the Schr\"{o}dinger equation%
\begin{equation}
\frac{d^{2}\psi (x)}{dx^{2}}+\frac{2m}{\hslash ^{2}}\left[ E-V\left(
x\right) \right] \psi (x)=0  \label{sch}
\end{equation}%
turns into the dimensionless form
\begin{equation}
\frac{d^{2}\psi (z)}{dz^{2}}-\left( \frac{z^{2}}{4}+a\right) \psi (z)=0
\label{airy}
\end{equation}%
where%
\begin{equation}
a=\frac{V(0)-E}{\hslash \omega }  \label{a}
\end{equation}%
The general solution for Eq. (\ref{airy}) can be written as a superposition
of definite-parity functions \cite{abr}%
\begin{equation}
\psi =c_{e}\,y_{1}(a,z)+c_{o}\,y_{2}(a,z)  \label{psi}
\end{equation}%
where%
\[
y_{1}(a,z)=e^{-z^{2}/4}\,M\left( \frac{a}{2}+\frac{1}{4},\frac{1}{2},\frac{%
z^{2}}{2}\right)
\]%
\begin{equation}
y_{2}(a,z)=e^{-z^{2}/4}\,z\,M\left( \frac{a}{2}+\frac{3}{4},\frac{3}{2},%
\frac{z^{2}}{2}\right)   \label{eq}
\end{equation}%
where $M(a,b,z)=\,_{1}F_{1}\left( a;b;z\right) $ is the confluent
hypergeometric function (Kummer's function)%
\begin{equation}
M(a,b,z)=\frac{\Gamma \left( b\right) }{\Gamma \left( a\right) }%
\sum_{n=0}^{\infty }\frac{\Gamma \left( a+n\right) }{\Gamma \left(
b+n\right) }\,\frac{z^{n}}{n!}  \label{m}
\end{equation}%
$\Gamma \left( z\right) $ is the gamma function, and $c_{e}$ and $c_{o}$ are
arbitrary constants. For $x>L$, the evanescent free-particle solution ($\psi
$ must vanish as $x\rightarrow \infty $) is expressed as
\begin{equation}
\psi =c\,e^{-kx}  \label{s2}
\end{equation}%
where $c$ is an arbitrary constant and
\begin{equation}
k=\sqrt{\frac{-2mE}{\hslash ^{2}}}=\frac{z_{L}^{2}}{L}\sqrt{\frac{r}{4}+%
\frac{a}{z_{L}^{2}}}  \label{epsilon}
\end{equation}%
Here,
\begin{equation}
z_{L}=\alpha L=\sqrt{2}\,\frac{\sqrt{\omega }L}{\sqrt{\hslash /m}}
\label{zl}
\end{equation}%
is the value of $z$ at $x=L$. Because $V\left( -x\right) =V\left( x\right) $%
, the Schr\"{o}dinger equation is invariant under space inversion ($%
x\rightarrow -x$) and so we can choose solutions with definite parities. The
even ($\psi _{e}$) and odd ($\psi _{o}$) parity eigenfunctions on the entire
$x$-axis can be written as%
\begin{eqnarray}
\psi _{e}\left( x\right)  &=&c_{e}\,e^{-\alpha ^{2}x^{2}/4}\,M\left( \frac{a%
}{2}+\frac{1}{4},\frac{1}{2},\frac{\alpha ^{2}x^{2}}{2}\right) \left[ \theta
\left( x+L\right) -\theta \left( x-L\right) \right]   \nonumber \\
&&  \nonumber \\
&&+\,c\,e^{-k|x|}\left[ \theta \left( x-L\right) +\theta \left( -x-L\right) %
\right]   \label{funpp}
\end{eqnarray}%
\begin{eqnarray}
\psi _{o}\left( x\right)  &=&c_{o}\,\alpha \,x\,e^{-\alpha
^{2}x^{2}/4}\,M\left( \frac{a}{2}+\frac{3}{4},\frac{3}{2},\frac{\alpha
^{2}x^{2}}{2}\right) \left[ \theta \left( x+L\right) -\theta \left(
x-L\right) \right]   \nonumber \\
&&  \nonumber \\
&&+\,c\,e^{-k|x|}\left[ \theta \left( x-L\right) -\theta \left( -x-L\right) %
\right]   \label{funii}
\end{eqnarray}%
The even parity solutions satisfy the homogeneous Neumann condition at the
origin ($d\psi (x)/dx|_{x=0}=0$) and the odd ones the homogeneous Dirichlet
condition ($\psi (0)=0$). In this circumstance it is enough to concentrate
our attention on the positive side of the $x$-axis and use the continuity of
$\psi (x)$ and $d\psi (x)/dx$ at $x=L$. Making use of the recurrence
formulas involving $y_{1}$ and $y_{2}$ defined in (\ref{eq}) \cite{abr}%
\[
\frac{dy_{1}(a,z)}{dz}+\frac{z}{2}\,y_{1}(a,z)=\left( a+\frac{1}{2}\right)
y_{2}(a+1,z)
\]%
\begin{equation}
\frac{dy_{2}(a,z)}{dz}+\frac{z}{2}\,y_{2}(a,z)=y_{1}(a+1,z)  \label{recrec}
\end{equation}%
one has as a result%
\begin{eqnarray}
\frac{d\psi }{dx} &=&\alpha e^{-z^{2}/4}\left\{ c_{e}\,z\left[ \left( a+%
\frac{1}{2}\right) M\left( \frac{a}{2}+\frac{5}{4},\frac{3}{2},\frac{z^{2}}{2%
}\right) -\frac{1}{2}M\left( \frac{a}{2}+\frac{1}{4},\frac{1}{2},\frac{z^{2}%
}{2}\right) \right] \right.   \nonumber \\
&&  \nonumber \\
&&+\left. c_{o}\left[ M\left( \frac{a}{2}+\frac{3}{4},\frac{1}{2},\frac{z^{2}%
}{2}\right) -\frac{z^{2}}{2}M\left( \frac{a}{2}+\frac{3}{4},\frac{3}{2},%
\frac{z^{2}}{2}\right) \right] \right\}   \label{difdif}
\end{eqnarray}%
The continuity of $\psi $ at $x=L$ says that%
\begin{equation}
c\,e^{-kL}
\end{equation}%
is equal to%
\begin{equation}
c_{e}\,e^{-z_{L}^{2}/4}\,M\left( \frac{a}{2}+\frac{1}{4},\frac{1}{2},\frac{%
z_{L}^{2}}{2}\right)   \label{p1}
\end{equation}%
for even parity solutions, and equal to
\begin{equation}
c_{o}\,e^{-z_{L}^{2}/4}z_{L}\,M\left( \frac{a}{2}+\frac{3}{4},\frac{3}{2},%
\frac{z_{L}^{2}}{2}\right)   \label{i1}
\end{equation}%
for odd parity solutions. Matching $d\psi /dx$ at $x=L$ makes%
\begin{equation}
-kc\,e^{-kL}
\end{equation}%
equal to%
\begin{equation}
c_{e}\,\alpha \,e^{-z_{L}^{2}/4}\,z_{L}\left[ \left( a+\frac{1}{2}\right)
M\left( \frac{a}{2}+\frac{5}{4},\frac{3}{2},\frac{z_{L}^{2}}{2}\right) -%
\frac{1}{2}M\left( \frac{a}{2}+\frac{1}{4},\frac{1}{2},\frac{z_{L}^{2}}{2}%
\right) \right]   \label{p2}
\end{equation}%
for even parity solutions, and equal to%
\begin{equation}
c_{o}\,\alpha \,e^{-z_{L}^{2}/4}\,\left[ M\left( \frac{a}{2}+\frac{3}{4},%
\frac{1}{2},\frac{z_{L}^{2}}{2}\right) -\frac{z_{L}^{2}}{2}M\left( \frac{a}{2%
}+\frac{3}{4},\frac{3}{2},\frac{z_{L}^{2}}{2}\right) \right]   \label{i2}
\end{equation}%
for odd parity solutions. Remembering the definition of $k$ from (\ref%
{epsilon}) and dividing (\ref{p2}) by (\ref{p1}), and (\ref{i2}) by (\ref{i1}%
), one finds the quantization condition%
\begin{equation}
f=g  \label{alpha2}
\end{equation}%
where%
\begin{equation}
f=\left\{
\begin{array}{c}
\frac{1}{2}-\left( a+\frac{1}{2}\right) \frac{M\left( \frac{a}{2}+\frac{5}{4}%
,\frac{3}{2},\frac{z_{L}^{2}}{2}\right) }{M\left( \frac{a}{2}+\frac{1}{4},%
\frac{1}{2},\frac{z_{L}^{2}}{2}\right) } \\
\\
\frac{1}{2}-\frac{1}{z_{L}^{2}}\,\frac{M\left( \frac{a}{2}+\frac{3}{4},\frac{%
1}{2},\frac{z_{L}^{2}}{2}\right) }{M\left( \frac{a}{2}+\frac{3}{4},\frac{3}{2%
},\frac{z_{L}^{2}}{2}\right) }%
\end{array}%
\begin{array}{c}
\textrm{for even parity solutions} \\
\\
\\
\\
\textrm{for odd parity solutions}%
\end{array}%
\right.   \label{g}
\end{equation}%
and%
\begin{equation}
g=\sqrt{\frac{r}{4}+\frac{a}{z_{L}^{2}}}  \label{f}
\end{equation}%
By solving the quantization condition for $a$ in the range%
\begin{equation}
-r\,\left( \frac{z_{L}}{2}\right) ^{2}<a<0  \label{res}
\end{equation}%
one obtains the possible energy levels for a particle trapped in the
potential well by inserting the allowed values of $a$ in (\ref{a}). Hence,%
\begin{equation}
E=V\left( 0\right) +|a|\,\hslash \omega   \label{e2}
\end{equation}%
Notice that $a$ only depends on the potential parameters via $r$ and $z_{L}$.

\section{Qualitative analysis}

A few qualitative results can be obtained with the aid of a plot of the
functions $f$ and $g$ on the same grid. Figure 2 shows the behaviour of $f$
against $|a|$ for two different values of $\sqrt{\omega }L$, and $g$ $\ $for
three different values of $r$. The eigenenergies are determined by the
intersections of the curves defined by $f$ with the square-root function
defined by $g$ ($0<g<\sqrt{r}/2$). Without ever solving the quantization
condition one is now apt to draw some conclusions about the localized
oscillator. It is instructive to note that this process for determining the
spectrum for the localized oscillator looks similar to that one for the
square potential. Notwithstanding, the zeros and poles of $f$ do not occur
at regular intervals as they do for $\tan \left( x\right) $ and $\cot \left(
x\right) $.

Seen as a function of $|a|$, $f$ presents branches of monotonically
increasing curves limited by vertical asymptotes due to the zeros of $%
M\left( a/2+1/4,1/2,z_{L}^{2}/2\right) $ and $M\left(
a/2+3/4,3/2,z_{L}^{2}/2\right) $. For large $\sqrt{\omega }L$ and small $|a|$%
, the abscissae of those asymptotes become approximately $n+1/2$, where $n$
is a nonnegative integer, and so do the zeros of $f$.

Since the confluent hypergeometric function goes to $1$ as $z_{L}\rightarrow
0$, one has that $f\rightarrow |a|$ for even parity solutions and $%
f\rightarrow -\infty $ for odd ones as $\sqrt{\omega }L\rightarrow 0$. Then,
because the square-root function vanishes for $|a|=r\left( z_{L}/2\right)
^{2}$ just one eigenenergy, that one associated with an even parity
eigenfunction with $|a|\simeq 0$, is allowed.

The number of possible bound states grows with $r\left( z_{L}/2\right) ^{2}$
but it is restricted by the value of $|a|$ which makes the square-root
function vanish. Therefore, the bound-states solutions constitute a finite
set of solutions if the potential parameters are finite.

The spectrum consists of energy levels associated with eigenfunctions of
alternate parities. The number of allowed bound states grows as the
potential parameters increase and there is at least one solution, no matter
how small \ the parameters are. All the eigenenergies, in the sense of $|a|$%
, tend asymptotically to the values $n+1/2$ as $\sqrt{\omega }L\rightarrow
\infty $ ($n=0,1,2,3,\ldots $). The energy levels tend to higher energies as
the parameter $r$ increases. As a function of $\sqrt{\omega }L$, the energy
level is a monotonous increasing function for $r\leq 1$ but enclosing the
oscillator with vertical walls ($r>1$) makes the energy level to reach a
maximum for some value of $\sqrt{\omega }L$.

\section{Approximate analytical calculations}

Asymptotic representations and limiting forms for the confluent
hypergeometric function allow us to obtain approximate analytical results
for the special cases of low-lying states and high-lying states.

The asymptotic expression which determines $M\left( a,b,z\right) $ for $%
a\rightarrow -\infty $ reads \cite{abr}%
\[
M(a,b,z)=\Gamma (b)e^{z/2}\left( \frac{1}{2}bz-az\right) ^{\frac{1}{4}-\frac{%
b}{2}}\pi ^{-1/2}
\]%
\begin{equation}
\times \cos \left[ \sqrt{2bz-4az}+\left( \frac{1}{4}-\frac{b}{2}\right) \pi %
\right] \left[ 1+\mathcal{O}\left( |b/2-a|^{-1/2}\right) \right] ,\quad
\textrm{for }z\in \mathbb{R}
\end{equation}%
Since $\Gamma \left( 3/2\right) =\Gamma \left( 1/2\right) /2$, it follows
that the quantization condition expressed by (\ref{alpha2}) takes the form%
\begin{equation}
kL-\frac{z_{L}^{2}}{2}\simeq \left\{
\begin{array}{c}
\left( \sqrt{|a|}z_{L}\right) \tan \left( \sqrt{|a|}z_{L}\right)  \\
\\
-\left( \sqrt{|a|}z_{L}\right) \cot \left( \sqrt{|a|}z_{L}\right)
\end{array}%
\begin{array}{c}
\textrm{for even parity solutions} \\
\\
\textrm{for odd parity solutions}%
\end{array}%
\right.   \label{QC3}
\end{equation}%
A further simplification occurs for $V(L)-V(0)<<|E|<<|V(0)|$, when $%
kL>>z_{L}^{2}/2$:%
\begin{equation}
kL\simeq \left\{
\begin{array}{c}
\left( \sqrt{|a|}z_{L}\right) \tan \left( \sqrt{|a|}z_{L}\right)  \\
\\
-\left( \sqrt{|a|}z_{L}\right) \cot \left( \sqrt{|a|}z_{L}\right)
\end{array}%
\begin{array}{c}
\textrm{for even parity solutions} \\
\\
\textrm{for odd parity solutions}%
\end{array}%
\right.
\end{equation}%
In this case the eigenfunction inside  the well turns into%
\begin{equation}
\psi \left( x\right) \simeq c_{e}\,\cos \left( \sqrt{|a|}\alpha x\right) +%
\frac{c_{o}}{\sqrt{|a|}}\,\sin \left( \sqrt{|a|}\alpha x\right)
\end{equation}%
Here we considered high-lying states in a  harmonic potential extending far
down ($r>>1$) and got the solutions for a square well potential. It means
that we may neglect any effects associated with the bottom of $V(x)$ as far
as high-lying states are concerned. It is instructive to note that the
condition $V(L)-V(0)<<|V(0)|$ makes the bottom of the potential look flat.

For small $z$, the hypergeometric function $M\left( a,b,z\right) $ goes like%
\begin{equation}
M\left( a,b,z\right) =1+\frac{a}{b}\,z+\frac{a\left( a+1\right) }{2b\left(
b+1\right) }\,z^{2}+\ldots  \label{z0}
\end{equation}%
Thus, the quantization condition for $z_{L}<<1$ turns into%
\begin{equation}
\begin{array}{c}
|a|+\left( |a|^{2}-\frac{r}{4}\right) z_{L}^{2}\simeq 0 \\
\\
\,1+\frac{|a|}{3}\,z_{L}^{2}\simeq 0%
\end{array}%
\begin{array}{c}
\textrm{for even parity solutions} \\
\\
\\
\textrm{for odd parity solutions}%
\end{array}%
\end{equation}%
Hence, just one root is allowed: $|a|\simeq 0$ for the even parity solution.
This quasi-null eigenenergy solution and its very delocalized eigenfunction
are valid for $\sqrt{\omega }L<<\sqrt{\hslash /m}$ when $V(0)\simeq 0$. In
this case the potential looks like a little ripple, a shallow well.

On the other hand, for large values of $|z|$ one has \cite{abr}%
\begin{equation}
\frac{M\left( a,b,z\right) }{\Gamma \left( b\right) }\simeq \frac{%
e^{z}\,z^{a-b}}{\Gamma \left( a\right) }\,,\quad \textrm{for Re }z>0
\label{ass}
\end{equation}%
In conjunction with the identity $\Gamma \left( z+1\right) =z\,\Gamma \left(
z\right) $ and with the fact that $\Gamma \left( z\right) $ has simple poles
at $z=-n$ $\ $with $n=0,1,2,3,\ldots $, the insertion of (\ref{ass}) into (%
\ref{g}) furnishes%
\begin{equation}
f\simeq \left\{
\begin{array}{c}
-1/2 \\
\\
{\textrm{undefined}}%
\end{array}%
\begin{array}{c}
\textrm{for }|a|\neq n+1/2 \\
\\
\textrm{for }|a|=n+1/2%
\end{array}%
\right.   \label{asy}
\end{equation}%
for both even and odd parity solutions. The singular behaviour of $f$ when $%
|a|=n+1/2$ is the reason that it undergoes infinite discontinuities at those
values of $|a|$, as can be grasped from Figure 2. It follows that, for
sufficiently large $z_{L}$, the square-root function can be expressed by%
\begin{equation}
g\simeq \frac{\sqrt{r}}{2}  \label{squareroot}
\end{equation}%
and the values
\begin{equation}
|a|\simeq \left\{
\begin{array}{c}
2n+1/2 \\
\\
2n+3/2%
\end{array}%
\begin{array}{c}
\textrm{for even parity solutions} \\
\\
\textrm{for odd parity solutions}%
\end{array}%
\right.   \label{qc2}
\end{equation}%
fulfill the quantization condition. Eq. (\ref{qc2}) represents a convenient
approximation as far as one considers the lowest values of $|a|$. As a
matter of fact, the intersections of the functions $f$ and $g$ occur just
slightly below the abscissae of the vertical asymptotes of $f$. Indeed, a
better approximation is obtained as $r$ grows. Nevertheless, for all the
values of $r$, the agreement improves as $z_{L}$ gets larger. In this
approximation, $M\left( a,b,z\right) $ reduces to a polynomial of degree $n$
in $z$ when $a=-n$. In particular, for $b=1/2$ and $b=3/2$ one has \cite{abr}%
\begin{eqnarray}
H_{2n}\left( x\right)  &=&\left( -1\right) ^{n}\frac{\left( 2n\right) !}{n!}%
M\left( -n,\frac{1}{2},x^{2}\right)   \nonumber \\
&&  \nonumber \\
H_{2n+1}\left( x\right)  &=&\left( -1\right) ^{n}\frac{\left( 2n+1\right) !}{%
n!}\,2x\,M\left( -n,\frac{3}{2},x^{2}\right)   \label{he}
\end{eqnarray}%
where $H_{n}\left( x\right) $ is the Hermite polynomial. Therefore, for $%
\sqrt{\omega }L>>\sqrt{\hslash /m}$ one gets the condensed form%
\begin{equation}
E_{n}\simeq V\left( 0\right) +\left( n+\frac{1}{2}\right) \,\hslash \omega
,\quad \psi _{n}\left( x\right) \simeq N_{n}\,e^{-\alpha
^{2}x^{2}/4}H_{n}\left( \frac{\alpha x}{\sqrt{2}}\right)
\end{equation}%
where $N_{n}$ is a normalization factor. The approximate results for $\sqrt{%
\omega }L>>\sqrt{\hslash /m}$ are expected to be exact in the limit $%
L\rightarrow \infty $ when the potential goes over to the full-space
harmonic oscillator. It is comforting to note that the particular values of $%
|a|$ obtained from the quantization condition are the same as those which
make the eigenfunction normalizable on the interval $\left( -\infty ,+\infty
\right) $. The harmonic oscillator approximation for $\sqrt{\omega }L$
finite, though, is only reasonable for the low-lying states, i.e. for energy
levels so near of the bottom of the potential that edge effects can be
neglected.

\section{Exact results}

The only remaining question is how to determinate exact results. With the
eigenfunctions on the whole line expressed by (\ref{funpp}) and (\ref{funii}%
), the problem resumes to find the eigenenergies. Although the quantization
condition has no closed form solutions in terms of simpler functions, the
numerical computation of the allowed values of $|a|$ can be done easily with
a root-finding procedure of a symbolic algebra program.

Figure 3 is a plot of the first low-lying energy levels, in the sense of $|a|
$, as a function of $\sqrt{\omega }L$. The bound-states solutions of the
localized and unconfined oscillator constitute a finite set of solutions.
The number of allowed bound states increases with $\sqrt{\omega }L$ and
there is at least one solution, no matter how small is $\sqrt{\omega }L$.
All the eigenvalues tend asymptotically to the values $n+1/2$ as $\sqrt{%
\omega }L\rightarrow \infty $ ($n=0,1,2,3,\ldots $). The energy levels tend
toward higher energies as the parameter $r$ increases, as can also be seen
in Figure 2. As a function of $\sqrt{\omega }L$, the energy level is a
monotonous increasing function for $r\leq 1$ but enclosing the oscillator
with a square well potential ($r>1$) makes the energy level reach a maximum
for some value of $\sqrt{\omega }L$.

Figure 4 shows the results for the ground-state eigenfunction against $x$
for $\sqrt{\omega }L=3/2\sqrt{\hslash /m}$ and $L$ equal to a Compton
wavelength. Included for comparison is the ground-state eigenfunction for
the full harmonic oscillator. The normalization $\int_{-\infty }^{+\infty
}dx\,|\psi |^{2}=1$ was done numerically. The eigenfunctions for $r=2$ ($%
|a|\simeq 0.520$) and $r=1/2$ ($|a|\simeq 0.416$) differ from that for the
full harmonic oscillator. The approximation does better for $r=2$, as it
does for the eigenvalue. In fact, the agreement is not bad even though we
have used $\sqrt{\omega }L\sim \sqrt{\hslash /m}$. Just as expected from the
above qualitative analysis, a more successful agreement for all the values
of $r$ should be obtained for $\sqrt{\omega }L>>\sqrt{\hslash /m}$.

\section{Conclusions}

We have assessed the bound-state solutions of the Schr\"{o}dinger equation
with a localized and unconfined harmonic potential well. We have derived the
energy eigenvalue equation and shown explicitly the eigenfunctions. We have
discussed the structure of the solutions of the eigenvalue equation. The
structure of the eigenfunctions has also been presented. The satisfactory
completion of this task has been alleviated by the use of graphical methods
and tabulated properties of the confluent hypergeometric functions. Finally,
the exact results have been presented.

As mentioned in the introduction of this work, the three-parameter
oscillator potential presents a richness of physics which might be relevant
for calculations in different fields of solid state physics, particularly in
electronics and computer components. Furthermore, it renders a sharp
contrast to the oscillator confined by infinite walls \cite{aul}-\cite{vaw}.

\bigskip

\bigskip

\bigskip

\bigskip

\noindent \textbf{Acknowledgments}

\noindent This work was supported in part by means of funds provided by
CAPES and CNPq.

\bigskip

\begin{figure}[th]
\begin{center}
\subfigure[$r>1$]{
\includegraphics[width=6.2cm]{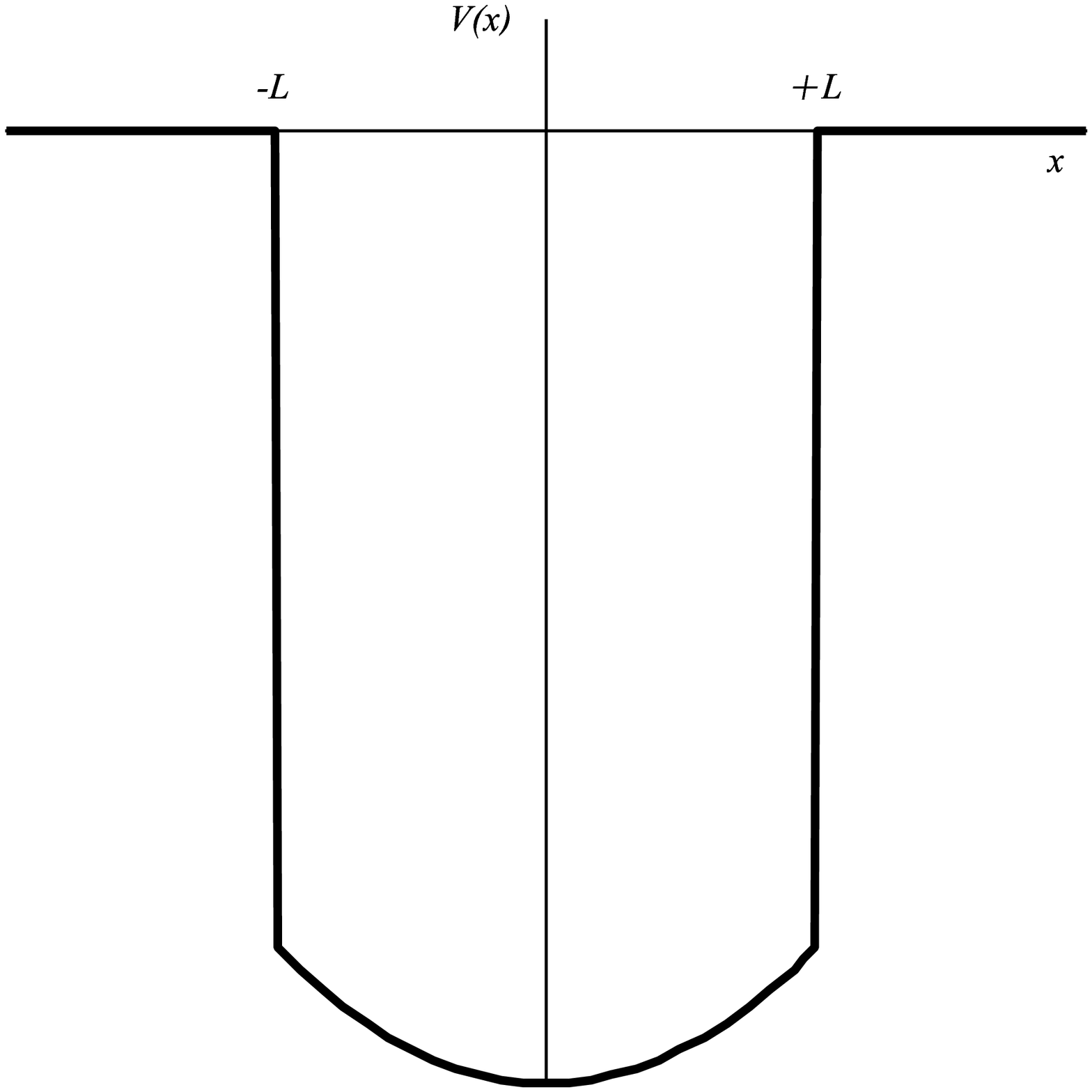}
\label{fig:Fig1a}
}
\subfigure[$r=1$]{
\includegraphics[width=6.2cm]{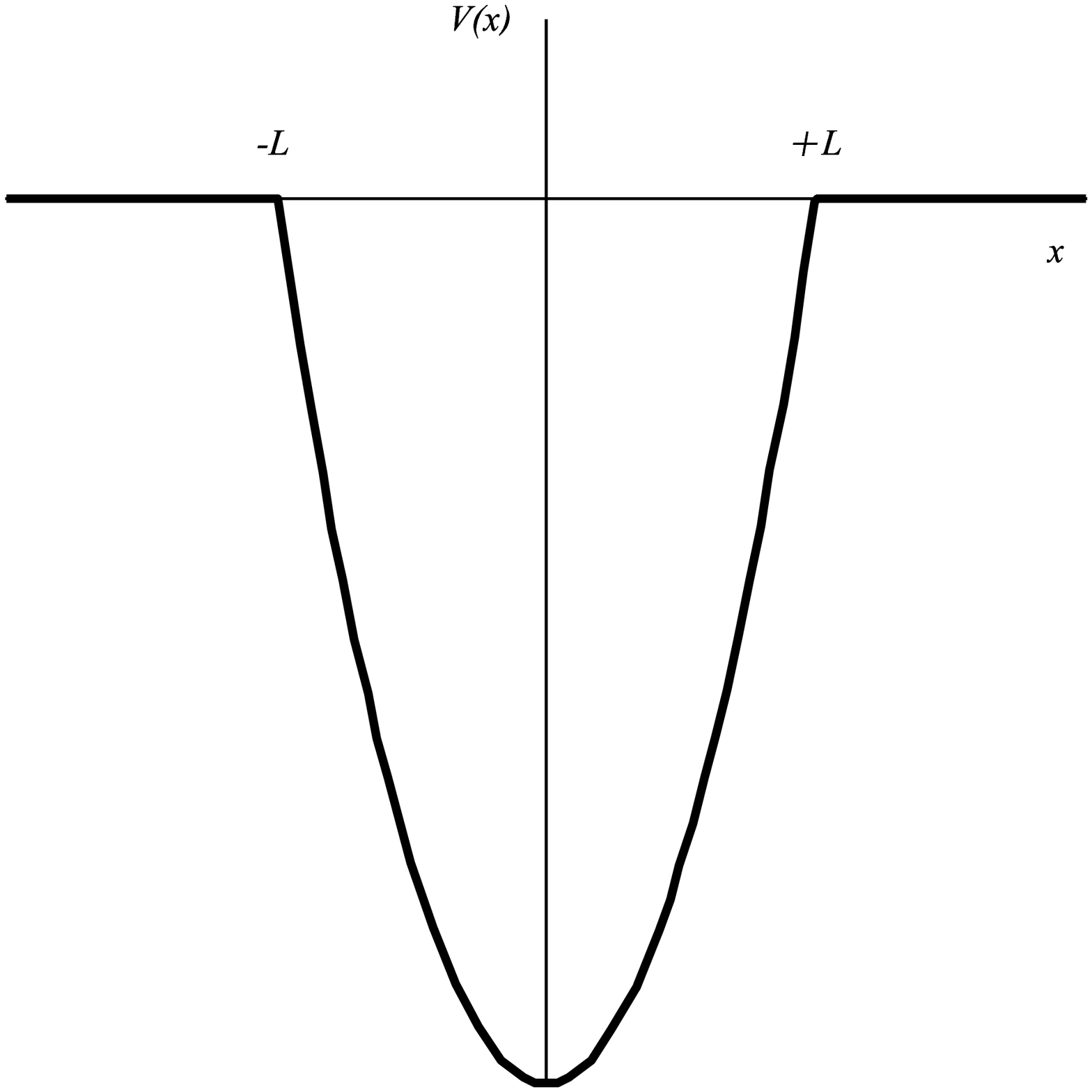}
\label{fig:Fig1b}
}
\subfigure[$0<r<1$]{
\includegraphics[width=6.2cm]{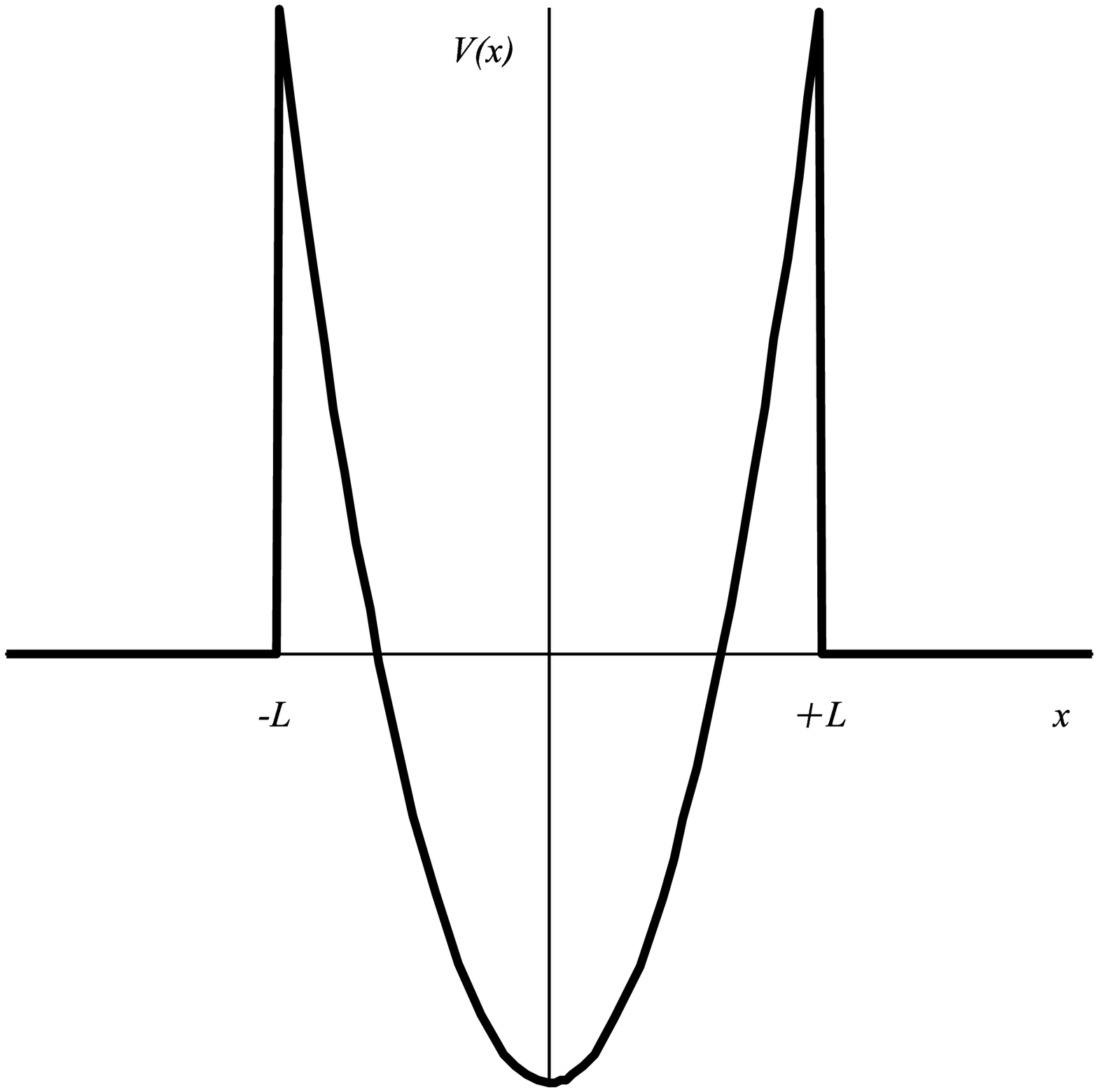}
\label{fig:Fig1c}
}
\subfigure[$r<0$]{
\includegraphics[width=6.2cm]{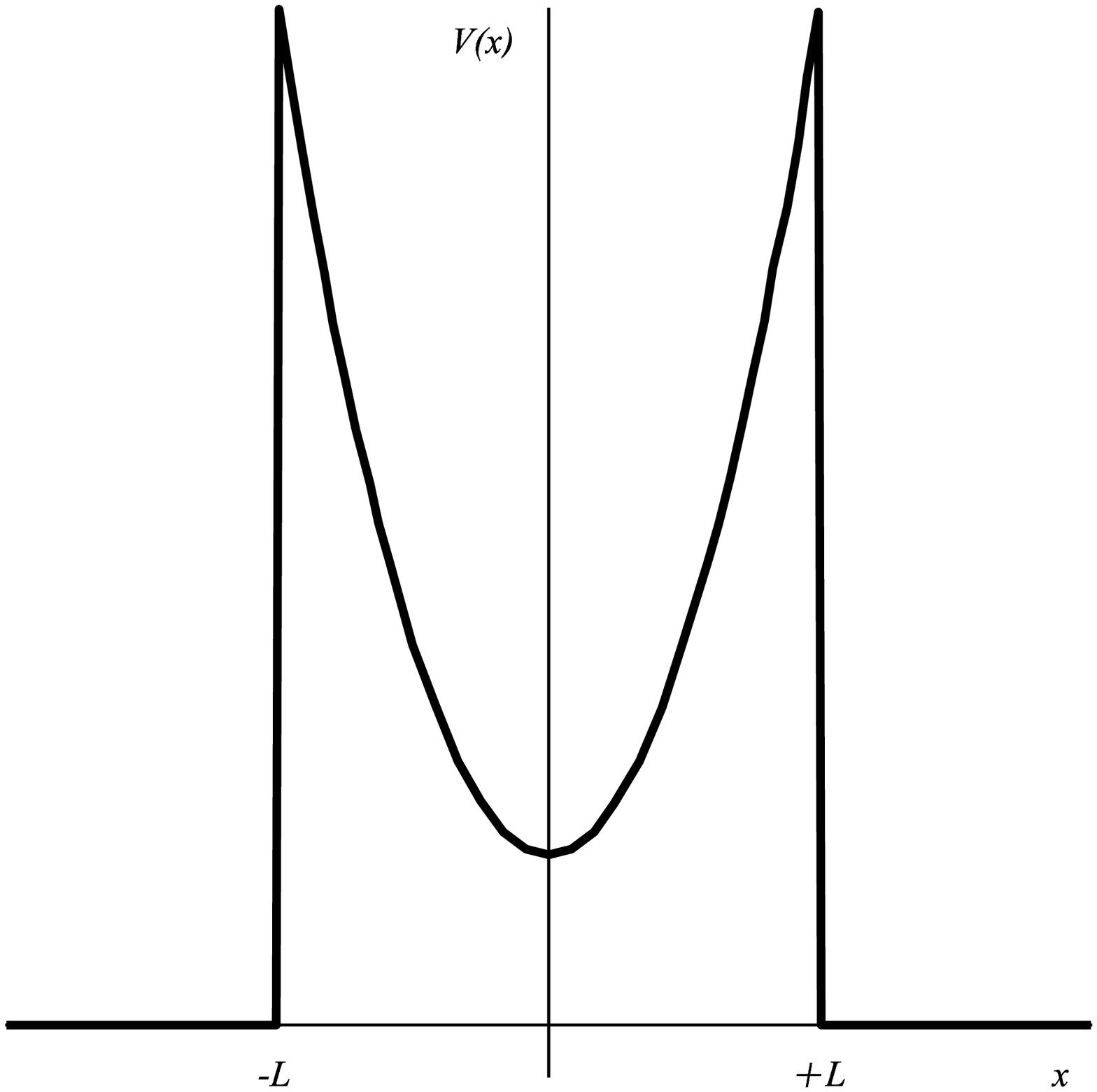}
\label{fig:Fig1d}
}
\end{center}
\caption{Profiles for $V(x)$.}
\end{figure}

\begin{figure}[th]
\begin{center}
\subfigure[$\sqrt{\omega }L=2\sqrt{\hslash /m}$]{
\includegraphics[width=10cm]{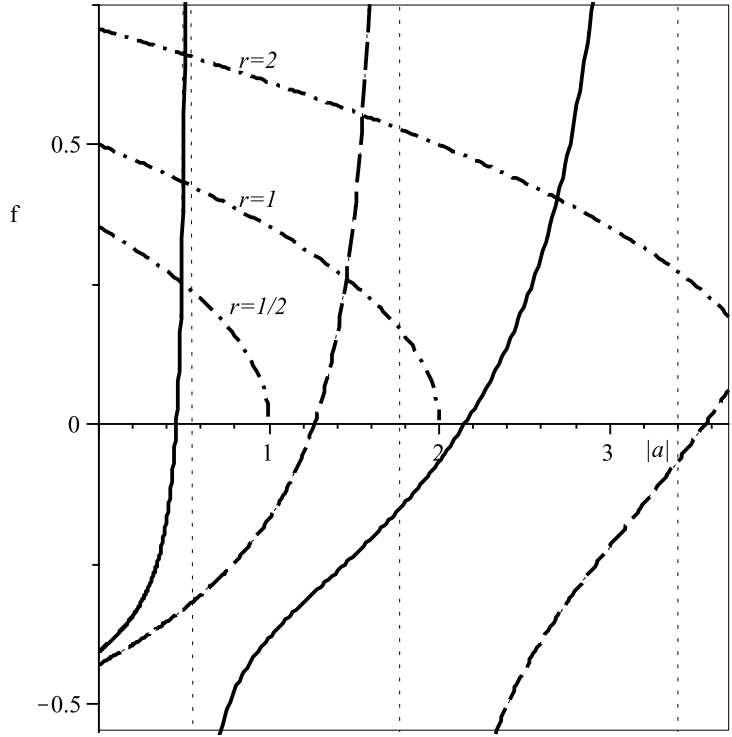}
\label{fig:Fig2a}
}
\subfigure[$\sqrt{\omega }L=3\sqrt{\hslash /m}$]{
\includegraphics[width=10cm]{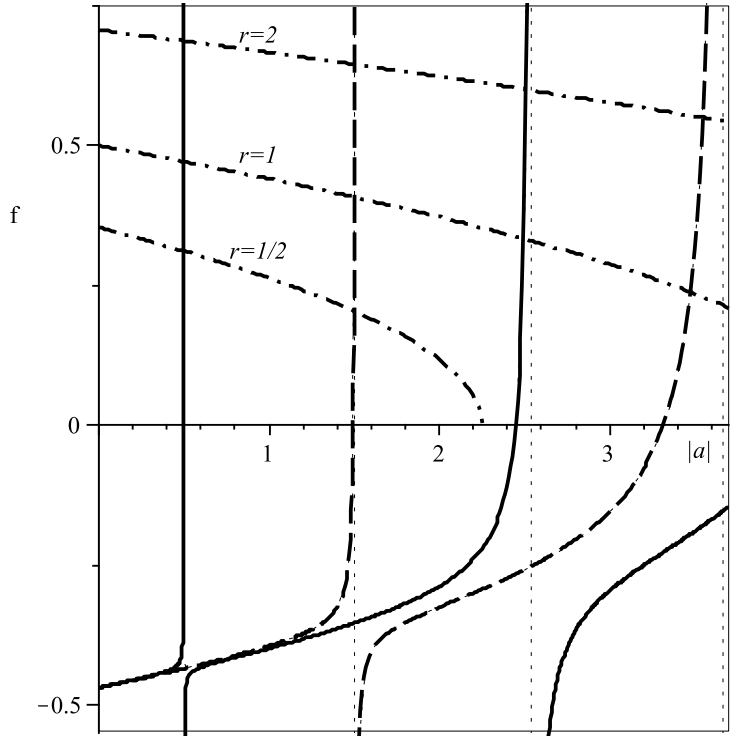}
\label{fig:Fig2b}
}
\end{center}
\caption{Graphical representation for $f$ as a function of $|a|$. The
continuous line for even parity solutions, the dashed line for odd ones, and
the dotted line for the asymptote. The dashed-dotted line stands for the
square-root function. }
\end{figure}

\begin{figure}[th]
\begin{center}
\includegraphics[width=10cm]{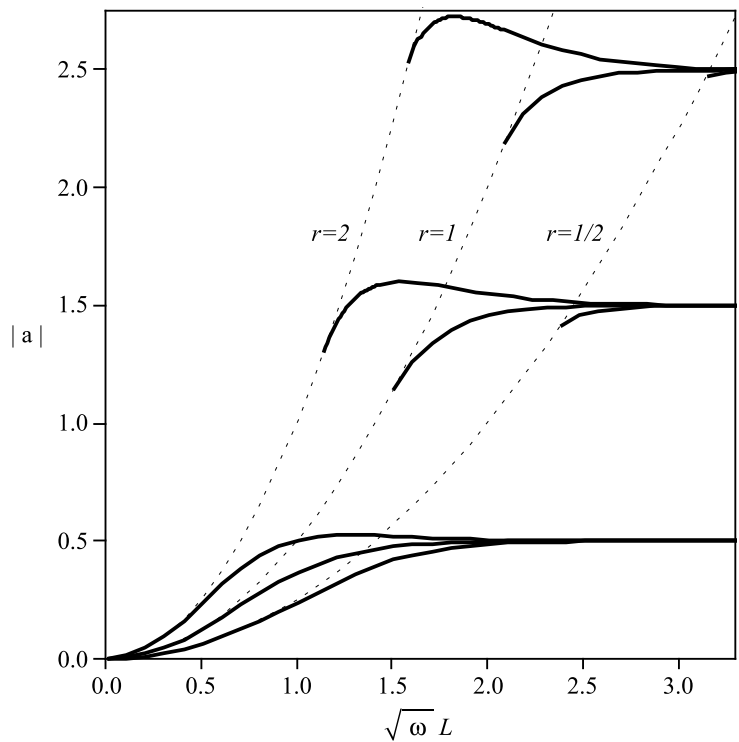} \label{fig:Fig3}
\end{center}
\par
\vspace*{-0.1cm}
\caption{$|a|$ for the first three energy levels as a function of $\protect%
\sqrt{\protect\omega }L$ (in units of $\protect\sqrt{\hslash /m}$) for three
representative values of $r$. The dotted parabola stands for the threshold
for the existence of bound states given by $|V(0)|/\hslash \protect\omega $.}
\end{figure}

\begin{figure}[th]
\begin{center}
\includegraphics[width=10cm]{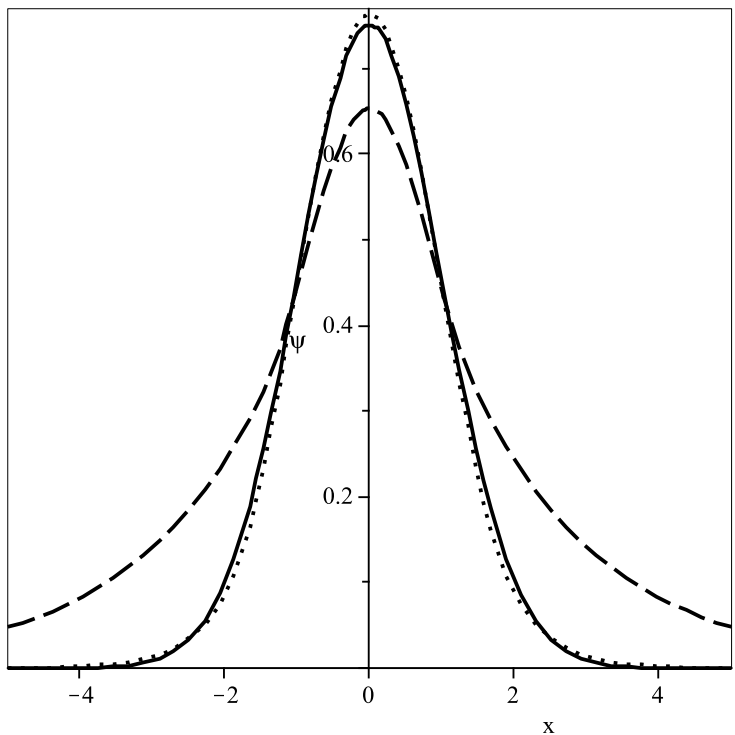} \label{fig:Fig4}
\end{center}
\par
\vspace*{-0.1cm}
\caption{Eigenfunction for the ground state as a function of $x$ for $%
\protect\sqrt{\protect\omega }L=3/2\protect\sqrt{\hslash /m}$ and $L$ equal
to the Compton wavelength. The continuous line for the full harmonic
oscillator, the dashed line for $r=1/2$ and the dotted line for $r=2$.}
\end{figure}

\end{document}